\documentclass[10pt,a4paper]{article}
\usepackage[utf8x]{inputenc}
\usepackage{ucs}
\usepackage{amsmath}
\usepackage{amsfonts}
\usepackage{amssymb}
\usepackage[color,final]{showkeys}
\usepackage{graphicx}
\usepackage[normalem]{ulem}
\usepackage[disable]{todonotes}
\usepackage{dsfont}
\usepackage[numbers]{natbib}
\usepackage[bookmarks=true]{hyperref}

\author{Jeroen Wouters$^{1}$ [jeroen.wouters@zmaw.de] and Valerio Lucarini$^{1,2}$\\ \\
$^{1}$Meteorologisches Institut, University of Hamburg\\KlimaCampus, Grindelberg 7,
20144 Hamburg, Germany\\
$^2$Department of Mathematics and Statistics\\University of Reading, Reading, RG6
6AX, UK}
\title{Multi-level Dynamical Systems: Connecting the Ruelle Response Theory and the Mori-Zwanzig Approach}

\newcommand{\mc}[1]{\mathcal{#1}}

\begin{document}
\maketitle

\begin{abstract}
In this paper we consider the problem of deriving approximate autonomous dynamics for a number of variables of a dynamical system, which are weakly coupled to the remaining variables. In a previous paper we have used the Ruelle response theory on such a weakly coupled system to construct a surrogate dynamics, such that the expectation value of any observable agrees, up to second order in the coupling strength, to its expectation evaluated on the full dynamics. We show here that such surrogate dynamics agree up to second order to an expansion of the Mori-Zwanzig projected dynamics. This implies that the parametrizations of unresolved processes suited for prediction and for the representation of long term statistical properties are closely related, if one takes into account, in addition to the widely adopted stochastic forcing, the often neglected memory effects.
\end{abstract}

\section{Introduction}

The investigation of multi-level systems is of primary interest for mathematics as well as for natural and social sciences, and is a central task of  complexity science. In multi-level systems it is possible to  separate the variables into at least two subsets, such that the variables within each subset are strongly coupled, while variables belonging to different subsets have a much weaker coupling. 

In most practical cases the dynamics of each level take place in distinct spatial and temporal scales, so that it is hard to define an optimal resolution for all the variables when we attempt to simulate the system or collect data \cite{fatkullin_computational_2004}.  Usually, one is interested in devising ways to account, at least approximately, for the impact of fast processes occurring at small spatial scales on slow variables, often describing large scale features, thus defining an effective autonomous dynamics for the slow variables.

If one assumes a vast time-scale separation between the slow variables $X$ and the fast variables $Y$, the averaging method \cite{arnold_hasselmanns_2001, kifer_recent_2004} allows for deriving a dynamics for the $X$ variables. Abramov~\cite{abramov_suppression_2011} has recently presented an application of this method to deriving a simplified dynamics for a system of geophysical relevance. Furthermore, projector operator techniques have been introduced in statistical mechanics with the goal of effectively removing the $Y$ variables. In particular, considerable interest has been raised by the Mori-Zwanzig approach, through which a formal - albeit practically inaccessible - solution for the evolution of the $X$ variables is derived \cite{zwanzig_nonequilibrium_2001}.

In a previous paper \cite{wouters_disentangling_2012} we have approached the problem of defining an accurate effective dynamics for the $X$ variable by focusing on the statistical properties of an observable $A(X)$ rather than on its trajectory. The starting point has been assuming that the coupling between the $X$ and $Y$ variables is weak, and treating such a coupling as a perturbation to the  autonomous dynamics of the $X$ and $Y$ variables, treated as unperturbed system. We have then adopted the response theory developed by Ruelle \cite{ruelle_differentiation_1997,ruelle_review_2009,lucarini08}, which allows to compute explicitly how the long-time averages of Axiom A dynamical systems change as a result of small perturbations to the flow  \cite{reick02,cessac07,majda07,lucarini09,lucarini_statistical_2011}.
The Axiom A class of systems, even if mathematically non-generic, is nevertheless assumed to serve as a good model for general physical systems, as is put forward by the chaotic hypothesis \cite{gallavotti_dynamical_1995,gallavotti_chaotic_1996}, which can be interpreted as an extension of the ergodic hypothesis to non-Hamiltonian systems \cite{gallavotti2006stationary}.
A detailed discussion of the relevance of the Axiom A dynamical systems for the description of actual physical processes is given in \cite{lucarini08,lucarini_statistical_2011}. Thus, we have derived explicit formulas to compute the changes in the expectation value of $A(X)$ up to the second order in the coupling strength between the $X$ and $Y$ variables. Moreover, we have been able to derive a surrogate perturbed dynamics for the system $X$ such that up to second order the expectation value of $A(X)$ is the same as that of the fully coupled $(X,Y)$ system, thus deriving an explicit parametrization of the coupling. The correction due to the coupling entail a deterministic contribution to the dynamics, a stochastic forcing expressed as a sum of multiplicative noise terms, and an integral expression which describes a memory term.

A question left open in \cite{wouters_disentangling_2012} was the link between the surrogate dynamics for the $X$ subsystem and the exact dynamics one would obtain by giving an explicit representation of the Mori-Zwanzig projection operator. Whereas we have shown that the proposed dynamics generate expectation values that are close to those of the fully coupled system, there may be many possible parametrizations that have this property. It is unclear to what extent the surrogate and full dynamics are related and whether they are close to each other. It is possible to construct a simplified model which reproduces well the statistical properties of the full model without describing well its time evolution. We anticipate that this paper gives the comforting answer that, indeed, the surrogate dynamics derived in \cite{wouters_disentangling_2012} and the dynamics projected using the  Mori-Zwanzig operator are identical up to second order of perturbation. As we will discuss in the conclusions, the fact that both the invariant measure and the dynamics of the full system are well approximated and the fact that a smoothing stochastic noise term is present, suggests the possibility of applying applying the fluctuation-dissipation theorem (FDT) for relating response and fluctuations of the X variables. This addresses some of the questions raised in \cite{ruelle_smooth_1999,ruelle_review_2009}, \cite{lucarini08,lucarini09}, \cite{lucarini_statistical_2011}, and \cite{colangeli_fluctuation-dissipation_2012}.

Our results rely on the presence of a relatively weak coupling, and not on the presence of a large time scale separation. Therefore, they may be of interest for a large number of complex systems. As a specific application, we wish to propose the derivation of parametrizations in geophysical fluid dynamics, where the simulation of long-term properties is the goal of climate modelling and the simulation of short-term behaviour is the problem of interest for weather forecast models. Geophysical fluid systems feature a very large range of time and space scales \cite{lucarini_modelling_2011,vallis_atmospheric_2006}, so parametrizations of unresolved small scales are necessary in numerical experiments \cite{orrell_model_2003,wilks_effects_2005,imkeller_stochastic_2001,hasselmann_stochastic_1976,arnold_hasselmanns_2001,saltzman_dynamical_2001}. Many of the approaches used up to now have been inspired by techniques coming from statistical mechanics, where a time scale separation between microscopic and macroscopic processes is often justifiable. In geophysical fluid systems, however, there is no such clear separation between time scales and hence the requirements for disregarding the memory term do not hold. Whereas deterministic and stochastic parametrizations are by now common in geophysical fluid dynamical models \cite{palmer_stochastic_2009}, memory effects are not incorporated. A final motivation for our current study of the connection between the long term statistics and dynamics is the increasing convergence between climate and weather prediction models. On the one hand there is the growing practice of benchmarking climate models by testing their prediction skill for weather forecast \cite{rodwell_palmer_2007}, while on the other hand there is an increasing interest in seamless prediction models \cite{palmer_2008}, that have a prediction range from days to years and longer.

This article is structured as follows. In Section \ref{sec:perturbation} we present the Dyson expansion for the evolution of the unperturbed and perturbed flows, and provide a way to treat in a unified way the Mori-Zwanzig and Ruelle's approaches. In particular, we present a formal derivation of the Ruelle's response formulas, which provides a simple way to derive the non-perturbative correction for the statistical properties of a general observable, and present the Mori-Zwanzig projection operator technique. In Section~\ref{sec:coupled} we then use this approach to deal with multi-level systems, thus deriving the explicit expression of the projected dynamics for the $X$ subsystem according to the Mori-Zwanzig formalism up to the second order of perturbation due to the coupling with the $Y$ subsystem. We show that such approximate dynamics agrees with what was obtained in \cite{wouters_disentangling_2012} using the Ruelle formalism. In Section \ref{sec:conclusions} we  present our conclusions and perspectives for future work.

\section{Expanding perturbed flows and averages}\label{sec:perturbation}
Given a dynamical system $\dot{x} = F(X)$, one can define a linear differential operator $L = (F.\nabla)$ describing the evolution of observables, i.e. $\dot{A}=LA$. Formally, the solution of $A$ over time is then given by $A(t)=\Pi(t)A(0)$ where $\Pi(t)=\exp(Lt)$, .
Both the Mori-Zwanzig projection operator technique~\cite{mori_transport_1965,zwanzig_memory_1961} and the response theory of natural invariant measures~\cite{ruelle_differentiation_1997} feature an expansion of evolution operators $\Pi(t)$. As described in \textit{e.g.} \cite{evans_statistical_2008}, these relations can be easily derived formally in the resolvent formalism, by taking the Laplace transform \todo{L unbounded\\ $s \notin \sigma(L)$} of such operator exponentials:
\begin{equation}
\mc{L}\lbrace \Pi \rbrace (s) = \int_0^\infty dt \exp(Lt) \exp(-ts) = (s - L)^{-1} \label{eq:laplace}
\end{equation}
If $L$ consists of a perturbation around an operator $L_0$, i.e. $L= L_0 + L_1$, with $L_{1}$ small in an appropriate sense, we can expand the Laplace transform using the equality
\begin{equation}
(A+B)^{-1} = A^{-1} - A^{-1} B (A+B)^{-1} \,.\label{expans}
\end{equation}
In the case of the Laplace transform in Eq.~\ref{eq:laplace}, we take $A=s-L_0$ and $B=-L_1$, so that the $A^{-1}$ and $(A + B)^{-1}$ terms are themselves Laplace transforms of $\Pi_0(t)=\exp(L_0 t)$ and $\Pi(t)$ respectively. Making use of a non-commutative version of the fact that the Laplace transform of a convolution is the product of the transform, this results in the following expansion of the evolution operator $\Pi(t)$:
\begin{align}
\Pi(t) = \Pi_0(t) + \int_0^t d\tau \Pi_0(t-\tau) L_1 \Pi (\tau) \label{eq:dyson}
\end{align}

Another expansion can be obtained when making use of the following equality for operator inverses:
\begin{equation}
(A+B)^{-1} = A^{-1} - (A+B)^{-1} B A^{-1} \,.
\end{equation}
This gives rise to the following decomposition of $\Pi(t)$:
\begin{align}
\Pi(t) = \Pi_0(t) + \int_0^t d\tau \Pi(t-\tau) L_1 \Pi_0 (\tau) \label{eq:dyson2}
\end{align}

\subsection{Projection operator techniques}
\label{sec:mz}
In the case of Mori-Zwanzig a projection is carried out on the level of the observables to remove unwanted, irrelevant and usually fast degrees of freedom. Here the expansion is performed around the evolution that involves only the relevant part of the phase space. If a dynamical system is defined on a Hilbert space $\mc{Z}$ with a relevant subspace $\mc{X}$ and its orthogonal complement $\mc{Y} = \mc{X}^\perp$, then one defines a projection $\mc{P}$ of functions on the full phase space to functions on the restricted phase space $\mc{X}$. For example, one can take a conditional expectation with respect to a measure on $\mc{Z}$:
\begin{align*}
(\mc{P} A) (x) = \frac{\int_{\mc{Y}} A(x,y) \rho(x,y) dy}{\int_{\mc{Y}} \rho(x,y) dy} \,.
\end{align*}
The derivation given by \citeauthor{zwanzig_nonequilibrium_2001} in Chapter 8 of \cite{zwanzig_nonequilibrium_2001} for a generalized Langevin equation is based on Eq. \ref{eq:dyson2}. We write the Liouville equation for an observable $A$ as
\begin{align*}
\frac{d A(t)}{dt} = LA(t) = e^{tL} LA = e^{tL} (\mc{P} + \mc{Q}) L A
\end{align*}
with $\mc{Q} = 1 - \mc{P}$.
The factor $exp(tL)$ in the term involving $\mc{Q}$ can be further expanded by making use of Eq. \ref{eq:dyson2} with $L_0 = \mc{Q}L$. This gives the following equation
\begin{align*}
\frac{d A(t)}{d t} = e^{tL} \mc{P}L A + ( e^{t\mc{Q}L} + \int_0^t ds \, e^{(t-s)L} \mc{P}L e^{s \mc{Q}L}) \mc{Q}LA
\end{align*}
In \cite{zwanzig_nonequilibrium_2001} it is argued that this equation is a generalization of the Langevin equation, where the second term is a correlated noise term dependent on the initial conditions of the irrelevant degrees of freedom and the third term represent the memory of the system due to the presence of irrelevant variables that have interacted with the relevant ones in the past. If there is a time scale difference between relevant and irrelevant variables the noise becomes white, while the memory term can in this case be neglected, as the irrelevant variables decorrelate quickly. 

\subsection{Response theory}
\label{sec:response}

The goal of response theory on the other hand is to describe changes in the averages of observables of dynamical systems over a long period of time:
\begin{align}
\tilde{\rho}(A) = \lim_{T \rightarrow \infty} \frac{1}{T} \int_0^T dt A(x(t)) = \lim_{T \rightarrow \infty} \frac{1}{T} \int_0^T dt \Pi(t) A(x(0)) \label{eq:srb}
\end{align}
We again assume that the operator $L$ that determines $\Pi$ consists of a perturbation around an unperturbed evolution: $L= L_0 + L_1$, meaning that we can make use of the expansions \ref{eq:dyson} and \ref{eq:dyson2} to expand $\Pi(t)$ around $\Pi_0(t)$. The perturbing operator $L_1$ can derive from an external forcing, or as we will later see from a coupling of internal degrees of freedom. The averages of an observable $A$ for the unperturbed evolution, corresponding to $L_0$ and $\Pi_0$, will be denoted by $\rho(A)$.

We present here a different derivation than the one in \cite{wouters_disentangling_2012}, where the response formula was derived by iterating Eq.~\ref{eq:dyson} to obtain the response terms at different orders in $L_{1}$. These terms can then be summed over all orders to obtain an expression for the full change in expectation value (as in \cite{ruelle_nonequilibrium_1998}). Here instead we will first derive the equation expressing the full change in expectation value, which can then be expanded in orders of the perturbation.

By inserting Eq.~\ref{eq:dyson2} into Eq.~\ref{eq:srb}, we have that
\begin{align}
\tilde{\rho}(A) &= \rho(A) + \lim_{T \rightarrow \infty} \frac{1}{T} \int_0^T dt \int_0^t d\tau \Pi(\tau) L_1 \Pi_0(t-\tau) A(x(0)) \\
			&= \rho(A) + \lim_{T \rightarrow \infty} \frac{1}{T} \int_0^T d\tau \int_0^{T-\tau} dt \Pi(\tau) L_1 \Pi_0(t) A(x(0)) \\
			&= \rho(A) + \tilde{\rho} \left(\int_0^\infty dt L_1 \Pi_0(t) A \right)
\end{align}
or
\begin{align}
\tilde{\rho}(A) &= \rho \left( \left(1 - \int_0^\infty dt L_1 \Pi_0(t) \right)^{-1} A \right)
\end{align}
By expanding the resolvent operator, one gets the response terms at different orders of $L_1$:
\begin{align}
\tilde{\rho}(A) = \rho(A) &+ \rho \left(\int_0^\infty dt_1 L_1 \Pi_0(t_1) A \right) \nonumber \\ 
&+ \rho \left(\int_0^\infty dt_1 \int_0^\infty dt_2 L_1 \Pi_0(t_1) L_1 \Pi_0(t_2) A \right) + \ldots \label{eq:response_all_order}
\end{align}
This expression is identical to those in \cite{ruelle_nonequilibrium_1998}.
\section{Coupled systems}\label{sec:coupled}

The response theory described in Section \ref{sec:response} is quite general, in the sense that we have not defined the form of the perturbing operator $L_1$. The perturbation can be a specified external forcing or an internal coupling of degrees of freedom. Here we choose the latter in order to make the comparison to the Mori-Zwanzig formalism. The dynamical system is given by an uncoupled vector field $F$ and a coupling function $\Psi$:
\begin{align}
\frac{dX}{dt} &= F_X(X) + \Psi_X(Y) \nonumber \\
\frac{dY}{dt} &= F_Y(Y) + \Psi_Y(X) \label{eq:coupled_dyn_syst}
\end{align}
Writing this in terms of observables, we have
\begin{align*}
\frac{dA(X,Y)}{dt} = & (L_X(X,Y) + L_Y(X,Y)) A(X,Y)  \\
= &(F_X(X) + \Psi_X(Y)).\nabla_X A(X,Y) \\
 & \; + (F_Y(Y) + \Psi_Y(X)).\nabla_Y A(X,Y)
\end{align*}
where $\nabla_X$ and $\nabla_Y$ denote the gradients with respect to the variables in $X$ and in $Y$ respectively, $L_X= (F_X + \Psi_X) \nabla_X$ and $L_Y= (F_Y + \Psi_Y) \nabla_Y$.

\subsection{Response theory}
The response of the unperturbed system to the coupling can be calculated by taking $L_0 = F_X(X).\nabla_X + F_Y(Y).\nabla_Y$ and $L_1 = \Psi_X(Y).\nabla_X + \Psi_Y(X).\nabla_Y$.

\begin{figure}
\centering
(a) \hspace{10pt} \includegraphics[scale=0.5]{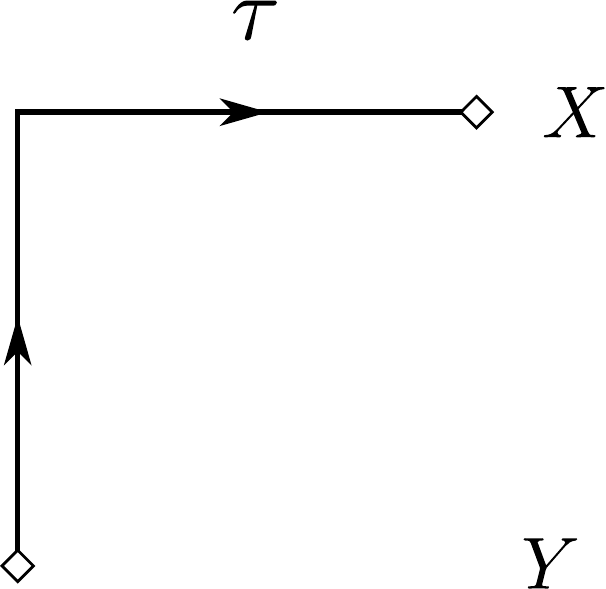} \hspace{1cm}
(b) \hspace{10pt} \includegraphics[scale=0.5]{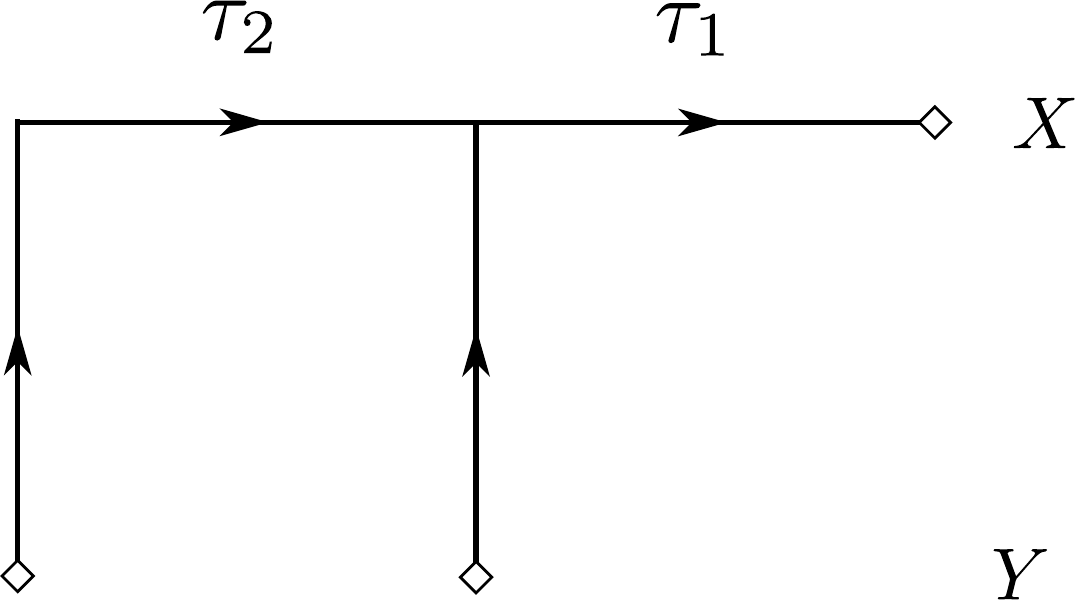}  \\ \vspace{1cm}
(c) \hspace{10pt} \includegraphics[scale=0.5]{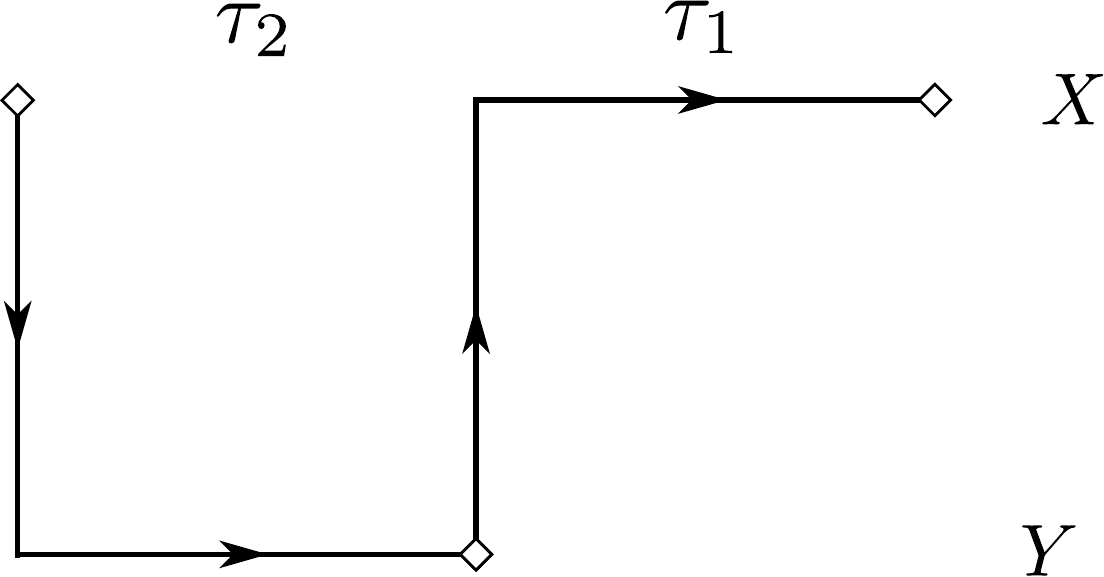}
\caption{Diagrams representing the response terms $\delta^{(1)} \rho(A_X|X,\tau)$ (diagram (a)), $\delta^{(2)} \rho(A_X|Y,\tau_2;X,\tau_1)$ (diagram (b)) and $\delta^{(2)} \rho(A_X|X,\tau_2;X,\tau_1)$ (diagram (c)).}
\label{fig:2nd_order2}
\end{figure}

By Eq.~\ref{eq:response_all_order}, the $n$-th order contribution to the response is calculated by integrating the impact of all $n$-time couplings over the times between the coupling interactions:
\begin{align}
\delta^{(n)}\rho(A) 
=& \int_{0}^{\infty} d\tau_1 \ldots \tau_{n} \sum_{\substack{i_1, \ldots , i_n \\ \in \lbrace X,Y\rbrace}} \delta^{(n)}\rho(A| i_1, \tau_1; \ldots ; i_n, \tau_n) 
\end{align}
where
\begin{align*}
\delta^{(n)}\rho(A| i_1, \tau_1; \ldots ; i_n, \tau_n) = \int \rho_0(dx) L_{1,i_1} \Pi_0(\tau_1) L_{1,i_2} \Pi_0(\tau_2) \ldots L_{1,i_n} \Pi_0(\tau_{n}) A(x)
\end{align*}
where $x=(X , Y)$ and $L_{1,i}$ represents an interaction affecting the $X$ or $Y$ subsystem, depending on the subscript:
\begin{align}
L_{1,X} &= \Psi_X(Y) \nabla_X \nonumber \\
L_{1,Y} &= \Psi_Y(X) \nabla_Y \label{eq:coupling_ops}
\end{align}
Any infinitesimal contribution to the response can hence be seen as a sequence of couplings that are activated subsequently ($i_1$ to $i_n$) and the times between the interactions ($\tau_1$ to $\tau_{n}$) as depicted in Figure~\ref{fig:2nd_order2}.

If one chooses as observable a function $A_X$ that is only dependent on $X$, all response contributions up to second order are $\delta^{(1)} \rho(A_X|X,\tau)$, $\delta^{(2)} \rho(A_X|Y,\tau_1;X,\tau_2)$ and $\delta^{(2)} \rho(A_X| X,\tau_1;X,\tau_2)$. The first order term $\delta^{(1)} \rho(A|X,\tau)$ is given by
\begin{align*}
\delta^{(1)} \rho(A|X,\tau) =& \int \rho_0(dx) L_X \Pi(\tau) A_X(x) \\
=& \rho_{0,Y}(\Psi_X(Y)) \rho_{0,X} ( \nabla_X A_X(f^{\tau}(X))) \,.
\end{align*}
The $\delta^{(2)} \rho(A_X|Y,\tau_1;X,\tau_2)$ and $\delta^{(2)} \rho(A_X| X,\tau_1;X,\tau_2)$ terms give
\begin{align*}
& \delta^{(2)} \rho(A_X|Y,\tau_1;X,\tau_2) \\
& \qquad  = \rho_{0,Y} \left( \nabla_Y \Psi_X ( f^{\tau_1} (Y) ) \right) \rho_{0,X} \left( \Psi_Y ( X) \nabla_X  (A_X \circ f^{\tau_2} ) ( f^{\tau_1}(X)) \right) \\
& \delta^{(2)} \rho(A_X|X,\tau_1;X,\tau_2) \\
& \qquad  = \rho_{0,Y} \left( \Psi_X ( Y )  \Psi_X ( f^{\tau_1} (Y) ) \right) \rho_{0,X} \left( \nabla_X ( \nabla_X ( A_X \circ f^\tau_2 ) (f^{\tau_1} (X)) \right)
\end{align*}
Note that since we perturb around the uncoupled system, the unperturbed measure $\rho_0$ is a product of invariant measures $\rho_{0,X}$ and $\rho_{0,Y}$ on the $X$ and $Y$ subsystems. For this reason and since the operators in \ref{eq:coupling_ops} are products of a multiplication and derivation operators that commute whenever the dependence is on different variables, each response term can be written as a product of a $\rho_{0,X}$ and $\rho_{0,Y}$ average.


As we have shown in \cite{wouters_disentangling_2012}, if one collects these first and second order responses to the coupling $\Psi$, an identical change in expectation values from the unperturbed $\rho_0$ can be obtained by adding a $Y$-independent perturbing operator $L_{1,p}$ to the uncoupled $L_0$. 
It was demonstrated that this can be accomplished by the following dynamical system: 
\begin{align}
\frac{dX(t)}{dt}&= F_X(X(t)) + M(X(t)) + \sigma_j (t) + \int_0^\infty d\tau h (\tau,X(t-\tau))\label{eq:second_model}
\end{align}
where $\sigma$ mimics the correlations present in the fluctuations of the coupling from its uncoupled mean:
\begin{align}
\langle \sigma_j(t) \sigma_l(t + \tau) \rangle &= \rho_Y ( \Psi^\prime_{X}(Y) \Psi^\prime_{X} (f^{\tau}_Y(Y))) \label{eq:model_correlation} \,,\\
\langle \sigma_j(t) \rangle &= 0 \nonumber \\
\Psi^\prime_{X}(Y) &= \Psi_{X}(Y) - \rho_Y(\Psi_{X}) \nonumber \,.
\end{align}
and $h$ is a kernel representing the memory effected by the presence of unresolved variables. Here $f_Y^\tau$ denotes the uncoupled evolution of $Y$ generated by $F_Y$.

The memory term $h$ is given by
\begin{align}
h(\tau,X) = \Psi_Y (X) \rho_Y( \nabla_Y \Psi_X (f_Y^\tau (Y)) ) \,. \label{eq:model_memory}
\end{align}

The $M$ term derives from $\delta^{(1)}\rho(A_X|X,\tau)$, the $\sigma$ term from $\delta^{(2)}\rho(A_X|X,\tau_1;X,\tau_2)$ and the $h$ term from $\delta^{(2)}\rho(A_X|Y,\tau_1;X,\tau_2)$.

It should be noted that the choice of parametrization is not unique. Any time-dependent forcing $\sigma$ with the correct two-point time-correlations will give the right response up to second order. Also for the memory term there is some freedom. One can also use 
\begin{align}
\frac{dX(t)}{dt}&= F_X(X(t)) + M(X(t)) + \sigma_j (t) + \int_0^\infty d\tau h (\tau,f_X^{(t-\tau)}(X_0)) \label{eq:second_model2}
\end{align}
as $X(t-\tau)$ is to zeroth order in $\Psi$ equal to $f_X^{(t-\tau)}(X_0)$. Hence the difference between the two parametrizations will be of order $\Psi^3$.

\subsection{Direct derivation of surrogate dynamics}
We now do a calculation in the style of the Mori-Zwanzig one in Section \ref{sec:mz} for the dynamical system given in Eq.~\ref{eq:coupled_dyn_syst} and for observables $A_X$ that only depend on the relevant variables $X$. As in Section~\ref{sec:mz}, we first do a projection of the evolution equation of $A_X$ to separate $X$ and $Y$ and then expand the evolution of $Y$.

The evolution equation for $A_X$ is given by
\begin{align}
(\frac{d}{dt} A_X) (X,Y,t) |_{t=0} & = (L_X A_X)(X,Y) = ((P L_X + Q L_X) A_X) (X,Y) \nonumber \\
	& = (F_X(X) + \rho_Y(\Psi_X) + (\Psi_X(Y)-\rho_Y(\Psi_X))) \nabla_X A_X(X) \label{eq:coupled_mz1}
\end{align}
where $P A(X,Y) = \int_\mc{Y} \rho_Y(dY) A(X,Y)$. We assume that $X$ and $Y$ start in $X_0$ and $Y_0$ at time $-t$. We want to find a formal solution for $\Psi_X(Y)$ that we can insert into the previous equation.

The evolution of $\Psi_X$ is given by
\begin{align*}
\Psi_X (Y) =& e^{t(L_X + L_Y)} \Psi_X (X_0,Y_0,-t) \\
			=& e^{t(L_X + L_Y)} \Psi_X (Y_0) 
\end{align*}
Making use of the decomposition of the Liouvillian $L=L_X + L_Y$ into $L_0(X_0,Y_0)= F_X(X_0) \nabla_X + F_Y(Y_0) \nabla_Y$ and $L_1(X_0,Y_0)=\Psi_X(Y_0) \nabla_X + \Psi_Y(X_0) \nabla_Y$, we get by repeated use of Eq.~\ref{eq:dyson} that
\begin{align}
\Psi_X(Y) =& e^{t(L_0 + L_1)} \Psi_X(Y_0) \nonumber \\
 =& e^{t L_0} \Psi_X(Y_0) + \int_0^t d \tau e^{(t-\tau)L_0} L_1 e^{\tau L_0} \Psi_X (Y_0) + O(L_1^2) \label{eq:coupled_mz_step2}
\end{align}
Inserting this equation in (\ref{eq:coupled_mz1}), we get
\begin{align}
&(\frac{d}{dt} A_X) (X,X_0,Y_0,t) |_{t=0} \nonumber \\
& \hspace{1cm} = \left(F_X(X) + \rho_Y(\Psi_X) + \tilde{\sigma}(t,Y_0) + \int_0^t d\tau \tilde{h}(t,\tau,X_{0},Y_0) \right) \nabla_X A_X(X)) \nonumber
\end{align}
where
\begin{align*}
\tilde{\sigma}(t,Y_0) =& e^{t F_Y(Y_0) \nabla_Y} \Psi_X(Y_0) - \rho_Y(\Psi_X) \\
\tilde{h}(t,\tau,X_0,Y_0) =& e^{(t-\tau)L_0(X_0,Y_0)} L_1(X_0,Y_0) e^{\tau L_0(X_0,Y_0)} \Psi_X (Y_0)
\end{align*}
Due to the commutation of $F_X \nabla_X$ and $F_Y \nabla_Y$, we have that
\begin{align*}
\tilde{h} =& e^{(t-\tau)(F_X \nabla_X + F_Y \nabla_Y)} (\Psi_X(Y_0) \nabla_X + \Psi_Y(X_0) \nabla_Y) e^{\tau F_Y \nabla_Y} \Psi_X (Y_0) \\
 =& \left( e^{(t-\tau)F_X \nabla_X}  \Psi_Y(X_0) \right) e^{(t-\tau)F_X \nabla_X} \nabla_Y e^{\tau F_Y \nabla_Y} \Psi_X (Y_0)
\end{align*}
If the coupled system is initialized in its stationary state, to zeroth order in $\Psi$, the $Y$ variable is distributed according to $\rho_Y$, the invariant measure under the flow generated by $F_Y$. The average of $\tilde{\sigma}$ is then zero and the auto-correlation is equal to that of $\sigma$ given before in Eq.~\ref{eq:model_correlation}:
\begin{align*}
\rho_Y (\tilde{\sigma}(t,Y_0)) &= 0 \\
\rho_Y \left(\tilde{\sigma}(t,Y_0) \tilde{\sigma}(t+\tau,Y_0) \right) &= \rho_Y \left( \Psi_X(Y_0)  e^{\tau F_Y \nabla_Y}  \Psi_X(Y_0) \right)
\end{align*}
and
\begin{align}
\rho_Y ( \tilde{h} ) = \Psi_Y(f^{t-\tau}_{X} (X_{0}))  \rho_Y \left( \nabla_Y e^{\tau F_Y \nabla_Y} \Psi_X (Y_0) \right) \label{eq:expected_memory}
\end{align}
This gives us the memory term of Eq.~\ref{eq:second_model2}.

\section{Summary and Conclusions}\label{sec:conclusions}

The first new result in this paper is a simple formal derivation of the Ruelle response theory \cite{ruelle_differentiation_1997,ruelle_review_2009,lucarini08} describing how the statistical properties of Axiom A systems are changed when the underlying dynamics is altered. We do not address issues of convergence, but simply present a simpler way of deriving the results, under the assumption that the integrals involved converge. We have shown that it is possible to obtain directly the exact expression for the expectation value of an observable $A$ computed according to the invariant probability measure of the altered dynamics as the expectation value of another suitably defined observable computed according to the invariant probability measure of the unperturbed dynamics. Such an expression can be expanded at all orders of perturbations, with the general term of order $n$ corresponding exactly to the $n^{th}$ order term obtained by Ruelle through perturbative expansion. Unsurprisingly, the exact result obtained through direct calculation agrees with what was derived by Ruelle through summation of the perturbative series \cite{ruelle_differentiation_1997}.

The Dyson expansion has been instrumental in approaching the problem of studying multi-level systems, by providing a peturbative expansion for the  Mori-Zwanzig projection operator \cite{zwanzig_nonequilibrium_2001}. Denoting by $X$ the subset of variables we are interested into and by $Y$ the subset of variables we want to project out, we have derived the effective projected dynamics describing the evolution of an observable of the $X$ variables only up to second order of perturbation. Such a dynamics is identical to the surrogate dynamics for the $X$ variables derived in \cite{wouters_disentangling_2012} by imposing, using the Ruelle response theory to describe the impact of the coupling of the $X$ and $Y$ variables, that the expectation value of any observable $A=A(X)$  evaluated on the invariant measure corresponding to the surrogate dynamics agrees up to second order of perturbation to its expectation value evaluated over the complete $(X,Y)$ system. It is important to note that, as discussed in \cite{wouters_disentangling_2012}, the surrogate dynamics is not unique, because we require agreement only up to second order. This result provides a connection between the Mori-Zwanzig and Ruelle formalisms, which are seemingly different, the first one pertaining to trajectories, the second one to expectation values: we have that if we are able to follow closely (on average) the individual trajectories, we are also able to model effectively the long-term statistical properties. Such a link between our ability to represent, in some sense, equally well, local and global properties in the phase space strongly relies on the fact that the projection operator leads to introducing a stochastic term and a memory term: the price we have to pay for neglecting the $Y$ degrees of freedom and still retaining a satisfactory representation of the $X$ dynamics on short and long time scales is going from a deterministic representation in terms of ordinary differential equations to an stochastic representation where integro-differential operators are involved. In particular, the consideration of memory effects marks the difference between what is discussed in this contribution and the classic method of averaging \cite{arnold_hasselmanns_2001, kifer_recent_2004}, which assumes that there is a vast time-scale separation between the two systems $X$ and $Y$, so that memory effects are negligible. Our approach, instead, relies on the presence of a relatively weak coupling between the $X$ and $Y$ variables. 

So far, we have been able to prove by direct calculation such a correspondence between the Mori-Zwanzig and Ruelle approaches only up to second order. but it is reasonable to conjecture that the same applies at any order $n$ of perturbation. If this is true, and taking the limit of $n\rightarrow \infty$, one would get that the exact Mori-Zwanzig projected dynamics provides the unique surrogate dynamics for the $X$ variables which is perfectly statistically compatible with the full $(X,Y)$ system for any observables of the $X$ variables only. Therefore, extending the proof we have given to all orders of perturbations would be extremely relevant. 

The results presented in this paper have relevance also in the context of the discussion on how to model practically and effectively high dimensional multiscale  systems. In particular, we refer to the problem of a) comparing models featuring different spatial resolutions; and b) constructing so-called parametrizations for the unresolved sub-scale processes. It is clear, from the rather general setting used here that increasing the resolution of a model amounts to enlarging the set of \textit{fast} variables $Y$ (this is particularly evident if one consider Galerkin-like expansions for the fields) coupled to the $X$ \textit{slow} variables, and devising parametrizations is nothing but approximating effectively the surrogated dynamics. This was partly discussed already in \cite{wouters_disentangling_2012} in the context of considering exclusively long-term statistical properties. What we have additionally learnt in this paper is that the Mori-Zwanzig projected dynamics, which is instead relevant for reproducing effectively the time evolution of the slow variables only, provides the surrogate dynamics we need to have a convincing statistics for the slow variables.

This has direct relevance for the modelling of geophysical fluid flows. It seems to support the idea of assessing the quality of climate models by testing their performance as tools for numerical weather prediction \cite{rodwell_palmer_2007}, and, more in general, points towards the direction of  the so-called seamless prediction \cite{palmer_2008}, which foresees the possibility of using the same models to perform forecasts over very different time scales, ranging from days to years and more. While usually the scholarly literature focuses on stochastic parametrizations as crucial tools in this direction \cite{palmer_stochastic_2009} , the present work underlines that the consideration of - usually neglected - memory effects is as important for achieving this goal.

We wish to make a final remark on the main findings of this paper. The right hand side of Eq. \ref{eq:second_model2} contains, as discussed above, a stochastic term. Therefore, the invariant measure of the dynamical system representing the surrogate dynamics is absolutely continuous with respect to Lebesgue. As discussed in \cite{ruelle_smooth_1999,ruelle_review_2009}, \cite{abramov_blended_2007}, \cite{lucarini08,lucarini09}, \cite{lucarini_statistical_2011}, this implies that the fluctuation-dissipation theorem can be applied, relating response and fluctuations. Since the invariant measure of the surrogate dynamics agrees, up to second order, with the projected measure of the X variables in the the full system and, as we have shown in this work, also the dynamics agrees closely to that of the full system, the response of the model should be close to that of the full system. The results therefore suggest the applicability of the fluctuation-dissipation theorem to the projected dynamics, as proposed using a different point of view in \cite{colangeli_fluctuation-dissipation_2012}. In particular, our results address the concerns expressed in \cite{lucarini_statistical_2011} on the applicability of the FDT in the context of climate dynamics and support the relatively positive findings obtained in this direction by \cite{langen_estimating_2005} \cite{gritsun_climate_2007}, \cite{abramov_information_2005} and \cite{cooper_climate_2011}.

\bibliographystyle{plainnat}
\bibliography{stochastic_modelling}

\begin{thebibliography}{38}
\providecommand{\natexlab}[1]{#1}
\providecommand{\url}[1]{\texttt{#1}}
\expandafter\ifx\csname urlstyle\endcsname\relax
  \providecommand{\doi}[1]{doi: #1}\else
  \providecommand{\doi}{doi: \begingroup \urlstyle{rm}\Url}\fi

\bibitem[Abramov(2012)]{abramov_suppression_2011}
R.~V. Abramov.
\newblock Suppression of chaos at slow variables by rapidly mixing fast
  dynamics through linear energy-preserving coupling.
\newblock \emph{Communications in Mathematical Sciences}, 10\penalty0
  (2):\penalty0 595--624, 2012.

\bibitem[Abramov and Majda(2008)]{majda07}
R.~V. Abramov and A.~Majda.
\newblock New approximations and tests of linear fluctuation-response for
  chaotic nonlinear forced-dissipative dynamical systems.
\newblock \emph{Journal of Nonlinear Science}, 18:\penalty0 303--341, 2008.
\newblock 10.1007/s00332-007-9011-9.

\bibitem[Abramov and Majda(2007)]{abramov_blended_2007}
R.~V. Abramov and A.~J. Majda.
\newblock Blended response algorithms for linear fluctuation-dissipation for
  complex nonlinear dynamical systems.
\newblock \emph{Nonlinearity}, 20\penalty0 (12):\penalty0 2793--2821, December
  2007.

\bibitem[Abramov et~al.(2005)Abramov, Majda, and
  Kleeman]{abramov_information_2005}
R.~V. Abramov, A.~Majda, and R.~Kleeman.
\newblock Information theory and predictability for low-frequency variability.
\newblock \emph{Journal of the Atmospheric Sciences}, 62\penalty0 (1):\penalty0
  65--87, January 2005.

\bibitem[Arnold(2001)]{arnold_hasselmanns_2001}
L.~Arnold.
\newblock Hasselmann's program revisited: The analysis of stochasticity in
  deterministic climate models.
\newblock \emph{Stochastic climate models}, 49:\penalty0 141--158, 2001.

\bibitem[Cessac and Sepulchre(2007)]{cessac07}
B.~Cessac and J.-A. Sepulchre.
\newblock Linear response, susceptibility and resonances in chaotic toy models.
\newblock \emph{Physica D: Nonlinear Phenomena}, 225\penalty0 (1):\penalty0 13
  -- 28, 2007.

\bibitem[Colangeli et~al.(2012)Colangeli, Rondoni, and
  Vulpiani]{colangeli_fluctuation-dissipation_2012}
M.~Colangeli, L.~Rondoni, and A.~Vulpiani.
\newblock Fluctuation-dissipation relation for chaotic non-hamiltonian systems.
\newblock \emph{Journal of Statistical Mechanics: Theory and Experiment},
  2012\penalty0 (04):\penalty0 L04002, April 2012.

\bibitem[Cooper and Haynes(2011)]{cooper_climate_2011}
F.~C. Cooper and P.~H. Haynes.
\newblock Climate sensitivity via a nonparametric {Fluctuation–Dissipation}
  theorem.
\newblock \emph{Journal of the Atmospheric Sciences}, 68\penalty0 (5):\penalty0
  937--953, May 2011.

\bibitem[Evans and Morriss(2008)]{evans_statistical_2008}
D.J. Evans and G.P. Morriss.
\newblock \emph{Statistical Mechanics of Nonequilibrium Liquids}.
\newblock Cambridge University Press, May 2008.

\bibitem[Fatkullin and {Vanden-Eijnden}(2004)]{fatkullin_computational_2004}
I.~Fatkullin and E.~{Vanden-Eijnden}.
\newblock A computational strategy for multiscale systems with applications to
  lorenz 96 model.
\newblock \emph{Journal of Computational Physics}, 200\penalty0 (2):\penalty0
  605--638, 2004.

\bibitem[Gallavotti(1996)]{gallavotti_chaotic_1996}
G.~Gallavotti.
\newblock Chaotic hypothesis: Onsager reciprocity and fluctuation-dissipation
  theorem.
\newblock \emph{Journal of Statistical Physics}, 84\penalty0 (5-6):\penalty0
  899--925, September 1996.

\bibitem[Gallavotti and Cohen(1995)]{gallavotti_dynamical_1995}
G.~Gallavotti and E.~G.~D. Cohen.
\newblock Dynamical ensembles in stationary states.
\newblock \emph{Journal of Statistical Physics}, 80\penalty0 (5-6):\penalty0
  931--970, September 1995.

\bibitem[Gallavotti and Roma(2006)]{gallavotti2006stationary}
G.~Gallavotti and F.~Roma.
\newblock Stationary nonequilibrium statistical mechanics.
\newblock \emph{Encyclopedia of Mathematical Physics, ed. JP Francoise, GL
  Naber, TS Tsun}, 3:\penalty0 530--539, 2006.

\bibitem[Gritsun and Branstator(2007)]{gritsun_climate_2007}
A.~Gritsun and G.~Branstator.
\newblock Climate response using a three-dimensional operator based on the
  {Fluctuation–Dissipation} theorem.
\newblock \emph{Journal of the Atmospheric Sciences}, 64\penalty0 (7):\penalty0
  2558--2575, July 2007.

\bibitem[Hasselmann(1976)]{hasselmann_stochastic_1976}
K.~Hasselmann.
\newblock Stochastic climate models, part {I}. theory.
\newblock \emph{Tellus}, 28\penalty0 (6):\penalty0 473--485, December 1976.

\bibitem[Imkeller and Storch(2001)]{imkeller_stochastic_2001}
P.~Imkeller and {J.-S.}~von Storch.
\newblock \emph{Stochastic Climate Models {(Progress} in Probability)}.
\newblock Birkh\"auser Basel, 1 edition, May 2001.

\bibitem[Kifer(2004)]{kifer_recent_2004}
Y.~Kifer.
\newblock Some recent advances in averaging.
\newblock \emph{Modern dynamical systems and applications: dedicated to Anatole
  Katok on his 60th birthday}, page 385, 2004.

\bibitem[Langen and Alexeev(2005)]{langen_estimating_2005}
P.~L. Langen and V.~A. Alexeev.
\newblock Estimating 2 × {CO2} warming in an aquaplanet {GCM} using the
  fluctuation-dissipation theorem.
\newblock \emph{Geophysical Research Letters}, 32\penalty0 (23):\penalty0
  L23708, December 2005.

\bibitem[Lucarini(2008)]{lucarini08}
V.~Lucarini.
\newblock Response theory for equilibrium and non-equilibrium statistical
  mechanics: Causality and generalized kramers-kronig relations.
\newblock \emph{Journal of Statistical Physics}, 131:\penalty0 543--558, 2008.
\newblock 10.1007/s10955-008-9498-y.

\bibitem[Lucarini(2009)]{lucarini09}
V.~Lucarini.
\newblock Evidence of dispersion relations for the nonlinear response of the
  {Lorenz} 63 system.
\newblock \emph{Journal of Statistical Physics}, 134:\penalty0 381--400, 2009.
\newblock 10.1007/s10955-008-9675-z.

\bibitem[Lucarini(2011)]{lucarini_modelling_2011}
V.~Lucarini.
\newblock Modelling complexity: the case of climate science.
\newblock \emph{arxiv:1106.1265}, June 2011.

\bibitem[Lucarini and Sarno(2011)]{lucarini_statistical_2011}
V.~Lucarini and S.~Sarno.
\newblock A statistical mechanical approach for the computation of the climatic
  response to general forcings.
\newblock \emph{Nonlinear Processes in Geophysics}, 18:\penalty0 7--28, 2011.

\bibitem[Mori(1965)]{mori_transport_1965}
H.~Mori.
\newblock Transport, collective motion, and {Brownian} motion.
\newblock \emph{Progress of Theoretical Physics}, 33\penalty0 (3):\penalty0
  423--455, March 1965.

\bibitem[Orrell(2003)]{orrell_model_2003}
D.~Orrell.
\newblock Model error and predictability over different timescales in the
  lorenz '96 systems.
\newblock \emph{Journal of the Atmospheric Sciences}, 60\penalty0
  (17):\penalty0 2219--2228, 2003.

\bibitem[Palmer and Williams(2009)]{palmer_stochastic_2009}
T.~N. Palmer and P.~Williams.
\newblock \emph{Stochastic Physics and Climate Modelling}.
\newblock Cambridge University Press, November 2009.

\bibitem[Palmer et~al.(2008)Palmer, J., A., and J.]{palmer_2008}
T.~N. Palmer, Doblas-Reyes~F. J., Weisheimer A., and Rodwell~M. J.
\newblock Toward seamless prediction: calibration of climate change projections
  using seasonal forecasts.
\newblock \emph{Bull. Amer. Meteor. Soc.,}, 89:\penalty0 459--470, 2008.

\bibitem[Reick(2002)]{reick02}
C.~H. Reick.
\newblock Linear response of the lorenz system.
\newblock \emph{Phys. Rev. E}, 66:\penalty0 036103, Sep 2002.

\bibitem[Rodwell and Palmer(2007)]{rodwell_palmer_2007}
M.~J. Rodwell and T.~N. Palmer.
\newblock Using numerical weather prediction to assess climate models.
\newblock \emph{Quarterly Journal of the Royal Meteorological Society},
  133\penalty0 (622):\penalty0 129--146, 2007.

\bibitem[Ruelle(1997)]{ruelle_differentiation_1997}
D.~Ruelle.
\newblock Differentiation of {SRB} states.
\newblock \emph{Communications in Mathematical Physics}, 187\penalty0
  (1):\penalty0 227--241, July 1997.

\bibitem[Ruelle(1998)]{ruelle_nonequilibrium_1998}
D.~Ruelle.
\newblock Nonequilibrium statistical mechanics near equilibrium: computing
  higher-order terms.
\newblock \emph{Nonlinearity}, 11\penalty0 (1):\penalty0 5--18, January 1998.

\bibitem[Ruelle(1999)]{ruelle_smooth_1999}
D.~Ruelle.
\newblock Smooth dynamics and new theoretical ideas in nonequilibrium
  statistical mechanics.
\newblock \emph{Journal of Statistical Physics}, 95\penalty0 (1):\penalty0
  393–468, 1999.

\bibitem[Ruelle(2009)]{ruelle_review_2009}
D.~Ruelle.
\newblock A review of linear response theory for general differentiable
  dynamical systems.
\newblock \emph{Nonlinearity}, 22\penalty0 (4):\penalty0 855--870, April 2009.

\bibitem[Saltzman(2001)]{saltzman_dynamical_2001}
B.~Saltzman.
\newblock \emph{Dynamical Paleoclimatology, Generalized Theory of Global
  Climate Change}.
\newblock Academic Press, 1st edition, November 2001.

\bibitem[Vallis(2006)]{vallis_atmospheric_2006}
G.~K. Vallis.
\newblock \emph{Atmospheric and Oceanic Fluid Dynamics: Fundamentals and
  Large-scale Circulation}.
\newblock Cambridge University Press, November 2006.

\bibitem[Wilks(2005)]{wilks_effects_2005}
D.~S Wilks.
\newblock Effects of stochastic parametrizations in the {Lorenz} '96 system.
\newblock \emph{Quarterly Journal of the Royal Meteorological Society},
  131\penalty0 (606):\penalty0 389--407, January 2005.

\bibitem[Wouters and Lucarini(2012)]{wouters_disentangling_2012}
J.~Wouters and V.~Lucarini.
\newblock Disentangling multi-level systems: averaging, correlations and
  memory.
\newblock \emph{Journal of Statistical Mechanics: Theory and Experiment},
  2012\penalty0 (03):\penalty0 P03003, March 2012.

\bibitem[Zwanzig(1961)]{zwanzig_memory_1961}
R.~Zwanzig.
\newblock Memory effects in irreversible thermodynamics.
\newblock \emph{Physical Review}, 124\penalty0 (4):\penalty0 983--992, 1961.

\bibitem[Zwanzig(2001)]{zwanzig_nonequilibrium_2001}
R.~Zwanzig.
\newblock \emph{Nonequilibrium statistical mechanics}.
\newblock Oxford University Press, 2001.

\end{thebibliography}

\end{document}